\documentclass[twocolumn,showpacs,preprintnumbers,amsmath,amssymb]{revtex4}

\usepackage{graphicx}
\usepackage{dcolumn}
\usepackage{bm}
\usepackage{color}
\usepackage{graphicx}
\usepackage{epstopdf}
\usepackage[hang,scriptsize]{subfigure}

\begin{document}


\title{Anchoring Bias in Online Voting}

\author{Zimo Yang}
\author{Zi-Ke Zhang}
\author{Tao Zhou}
\email{zhutou@ustc.edu}
\affiliation{Web Sciences Center, University of Electronic Science and Technology of China, Chengdu 610054, People's Republic of China}

\begin{abstract}

Voting online with explicit ratings could largely reflect people's preferences and objects' qualities, but ratings are always irrational, because they may be affected by many unpredictable factors like mood, weather, as well as other people's votes. By analyzing two real systems, this paper reveals a systematic bias embedding in the individual decision-making processes, namely people tend to give a low rating after a low rating, as well as a high rating following a high rating. This so-called \emph{anchoring bias} is validated via extensive comparisons with null models, and numerically speaking, the extent of bias decays with interval voting number in a logarithmic form. Our findings could be applied in the design of recommender systems and considered as important complementary materials to previous knowledge about anchoring effects on financial trades, performance judgements, auctions, and so on.

\end{abstract}


\maketitle

Uncovering human behavioral patterns, such as bursty nature of temporal activity \cite{Barabasi2005,Vazquez2006}, scaling laws of human travel \cite{Brockman2006,Gonzalez2008}, different selecting patterns of different kinds of users \cite{Shang2010,Zhang2012} onto different kinds of objects \cite{Vig2011,Chen2012}, is significant to understand many socioeconomic phenomena and provide high-quality services. Here we investigate online voting, which contains huge business value in e-commerce. Take recommender systems as an example, via analyzing online votes, they can automatically find out suitable products for every customer \cite{Lu}. So it could largely improve the performance of recommender systems, if we make clear the knowledge about how people vote \cite{Koren2010,Koren2011}.

In some systems, votes are confined to only two extremes--like or dislike, while in some other systems, people can vote with explicit ratings--usually from one star to five stars. Explicit ratings, however, do NOT signify rational judgments. Indeed, people's votes may be largely affected by prior votes \cite{Alon2012} and social pressure (like suggestions from friends) \cite{Huang2012}. We do not consider the aforementioned biases in this paper, in that when one votes on the systems we analyze here, neither others' votes nor social network services are provided for users, but by comparing with null models, considerable voting bias are still observed, which originates from internal decision-making processes of individuals, that is to say, people strongly tend to give a low rating again after voting on a low-quality object, as well as to give a high rating again after voting on a high-quality object. We name it as anchoring bias, since it is similar to the well-known anchoring effects in purchases \cite{Simonson2004,Wu2012}, auctions \cite{Ku2006,Beggs2009}, judgements \cite{Hunt1941,Thorsteinson2008} and estimations \cite{Kaustia2008,Cen2010} (see also the review article \cite{Furnham2011} and the references therein). Previous experiments \cite{Furnham2011} showed that even a randomly assigned initial value of an object could remarkably affect our estimation on its real value. This paper indicates that a prior vote on another object could affect our current vote because we may take that prior vote as an anchor. 

\begin{table}[htbp]
\centering
\caption{\label{Table I}Basic statistics of MovieLens and WikiLens. $N$, $M$ and $V$ denote the number of users, objects and ratings, respectively, $\rho=\frac{V}{NM}$ denotes the sparsity of the data, and $\langle r \rangle$ is the average rating over all votes.}
\begin{tabular}{cccccc}
\hline
\hline
Data Set & $N$ & $M$ & $V$ & $\rho$ & $\langle r \rangle$ \\
\hline
MovieLens & 6040 & 3952 & 1000292 & 0.042 & 3.58\\
WikiLens & 289 & 4951 & 26937 & 0.019 & 3.71\\
\hline
\hline
\end{tabular}
\end{table}

In this paper, we consider two real data sets, MovieLens and WikiLens. MovieLens is a movie rating systems with five stars (i.e., ratings can be 1, 2, 3, 4 and 5). The WikiLens is a generalized collaborative recommender system that allows its community to define object types (e.g., beer) and categories (e.g., microbrews), and vote on objects. Ratings in WikiLens can be $1, 1.5, \cdots, 4.5, 5$. Both the two data sets can be found in GroupLens research web (http://grouplens.org/), and their basic statistics are summarized in Table I.

A recommender system with explicit ratings can be described by a weighted bipartite network where each vote is represented by an edge connecting the corresponding user and object, and its weight is defined as the corresponding rating. The degree of a user is defined as the number of objects she has voted, while the degree of an object is the number of users who have voted on it. Figure 1 reports the degree distributions for users, which do not follow neither the power-law form nor the exponential form. In fact, they lie in between exponential and power-law forms, and can be well fitted by the so-called stretched exponential distributions \cite{Laherrere1998,Zhou2007}
\begin{equation}
p(k)\sim k^{c-1}\exp\left[-\left(\frac{k}{K}\right)^c\right],
\end{equation}
where $K$ is a constant and $c$ $(0<c<1)$ is the characteristic exponent. The borderline $c=1$ corresponds to the usual exponential distribution. For $c$ smaller than 1, the distribution presents a clear curvature in a log-log plot.

\begin{figure}
\includegraphics[width=1.6in]{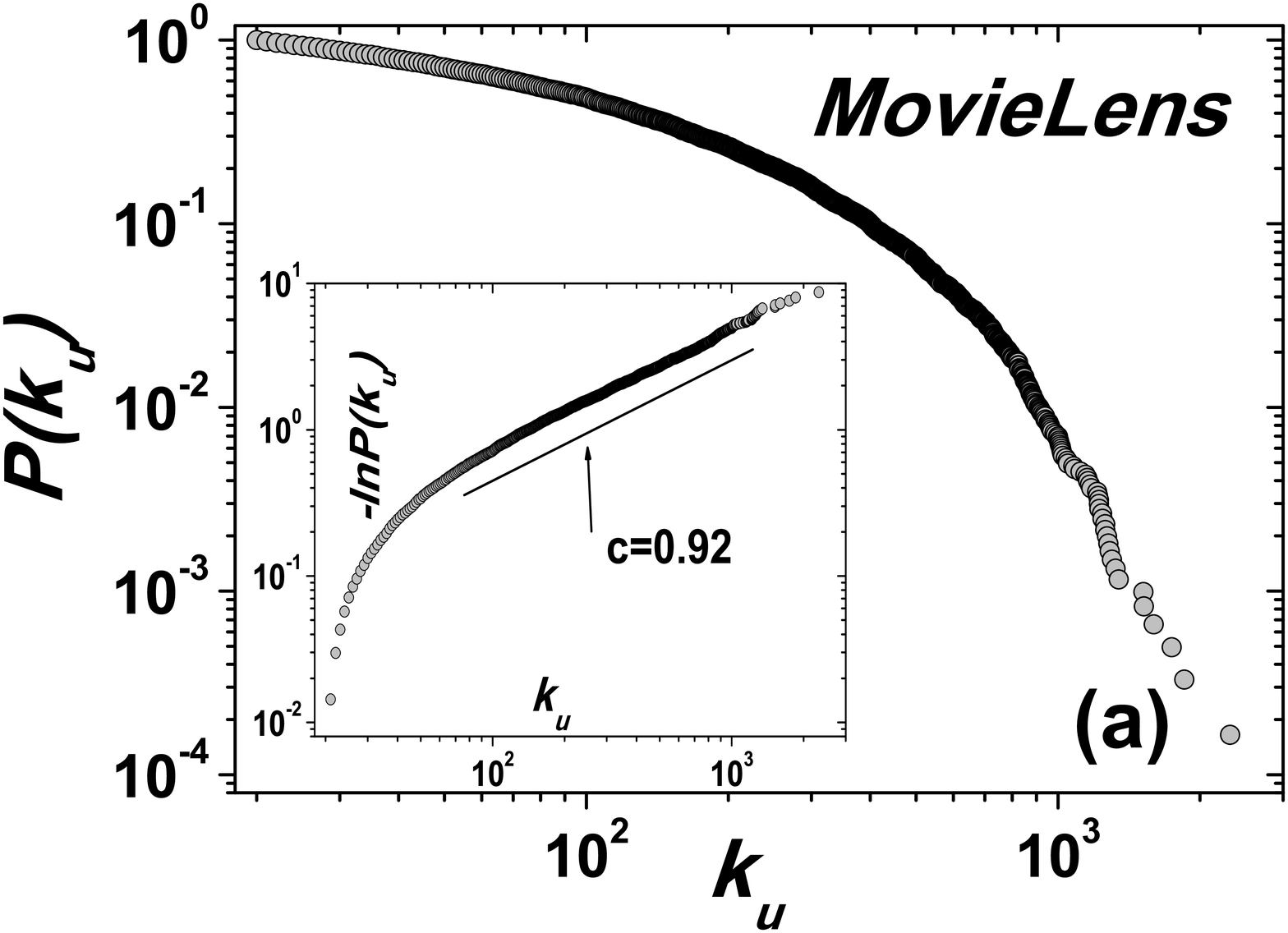}
\includegraphics[width=1.6in]{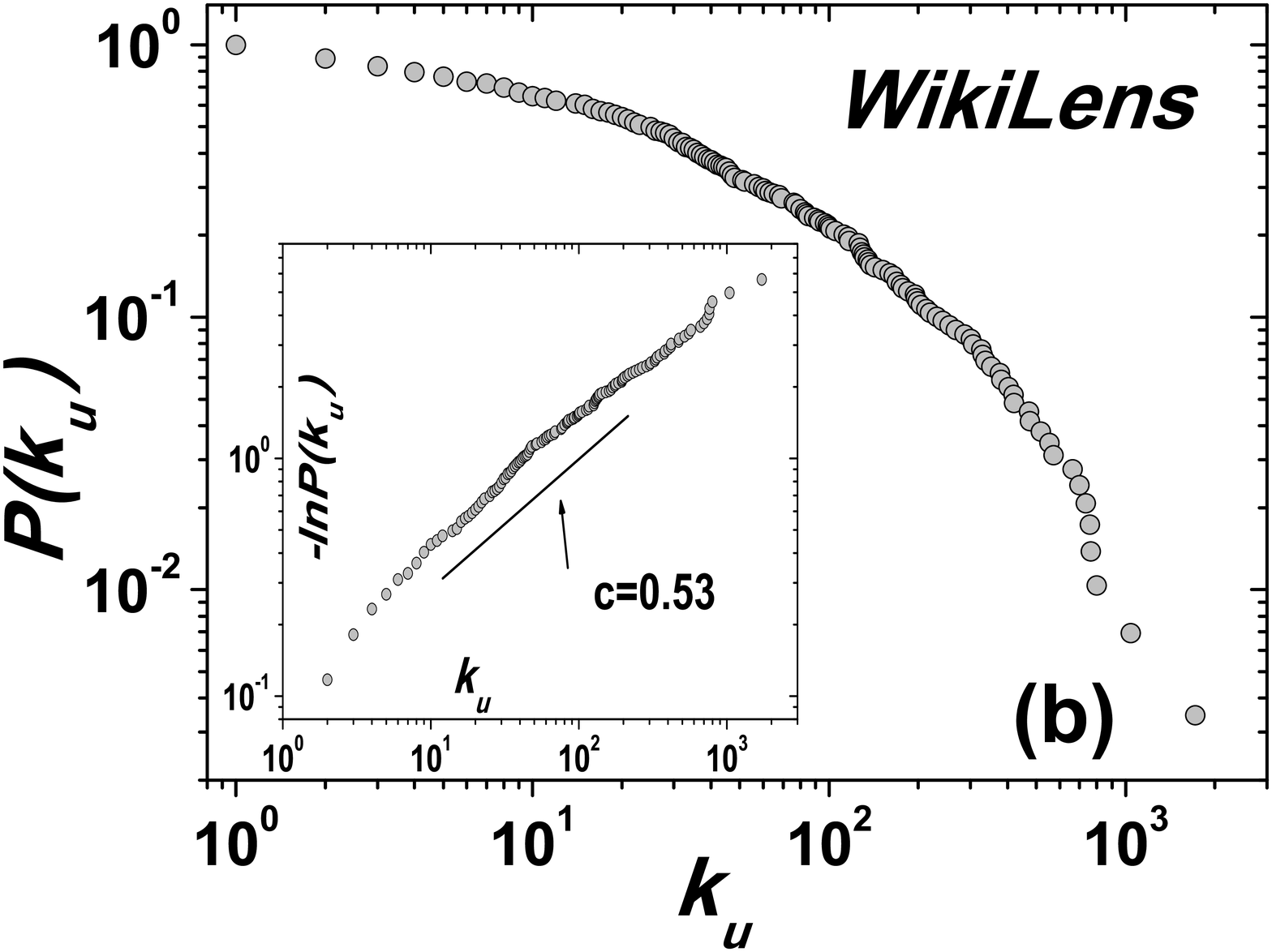}
\caption{Distributions of user degrees, which obey the stretched exponential form. We therefore plot the cumulative distribution $P(k_u)$ instead of $p(k_u)$ and show the linear fittings of $\ln(-\ln P(k_u))$ vs. $\ln k_u$ in the insets.}
\end{figure}

The exponent $c$ can be determined by considering the cumulative distribution
\begin{equation}
P(k)\sim\exp\left[-\left(\frac{k}{K}\right)^c\right],
\end{equation}
which can be rewritten as
\begin{equation}
\ln(-\ln P(k))\sim c\ln k.
\end{equation}
Therefore, using $\ln k$ as $x$-axis and $\ln(-\ln P(k))$ as $y$-axis, if the corresponding curve can be well fitted by a straight line, then the slope equals $c$. Accordingly, as shown in figure 1, the exponents $c$ for MovieLens and WikiLens are 0.92 and 0.53, respectively. Note that, the user degree distribution of MovieLens is very close to a usual exponential form. This kind of distributions often displays a mixture of power-law and exponential form \cite{Clauset2009}, and are usually fitted by stretched exponential function \cite{Laherrere1998}, power law with exponential cutoff \cite{Newman2001} or Mandelbrot law \cite{Ren2012}. Often, the head is closer to a power law while the tail shows exponential decay due to the limitation of people's ability in accessing information. Since the MovieLens data only consists of users having voted on no less than 20 movies, the head part cannot be observed and thus the distribution is close to an exponential form. Figure 2 reports the object degree distributions that also obey the stretched exponential form. Note that, in some other online user-object bipartite networks (e.g., audioscrobbler.com and delicious.com) where the number of users is huge and users are not required to vote on objects, the object degree can be very well characterized by power-law distribution \cite{Shang2010,Lambiotte2012}. The distinct statistics of object-degree distributions of the present systems have refined our knowledge about online user-object bipartite networks and raised open question about whether the huge number of users and/or less efforts of actions are necessary to the appearance of power laws.

\begin{figure}
\includegraphics[width=1.6in]{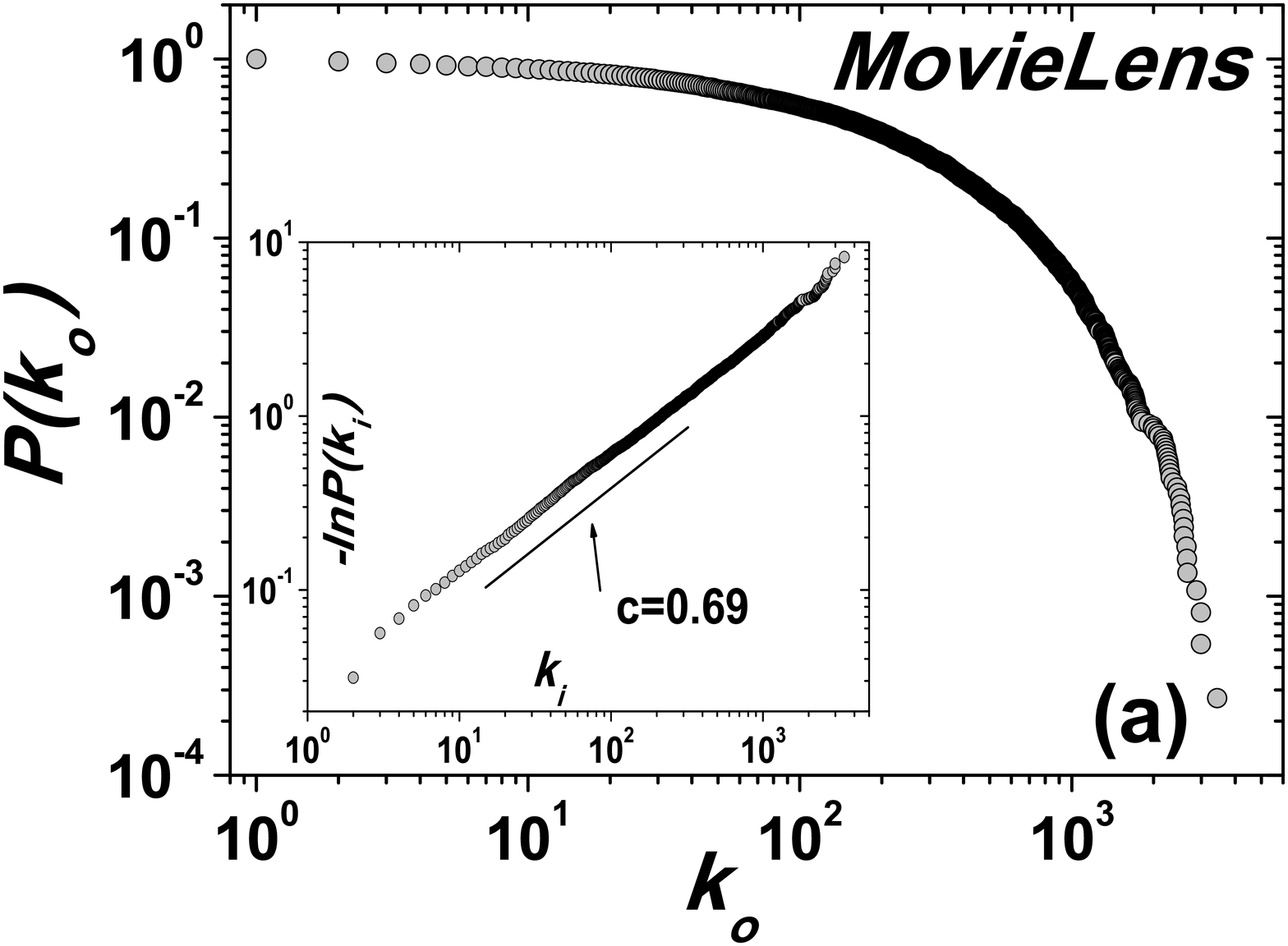}
\includegraphics[width=1.6in]{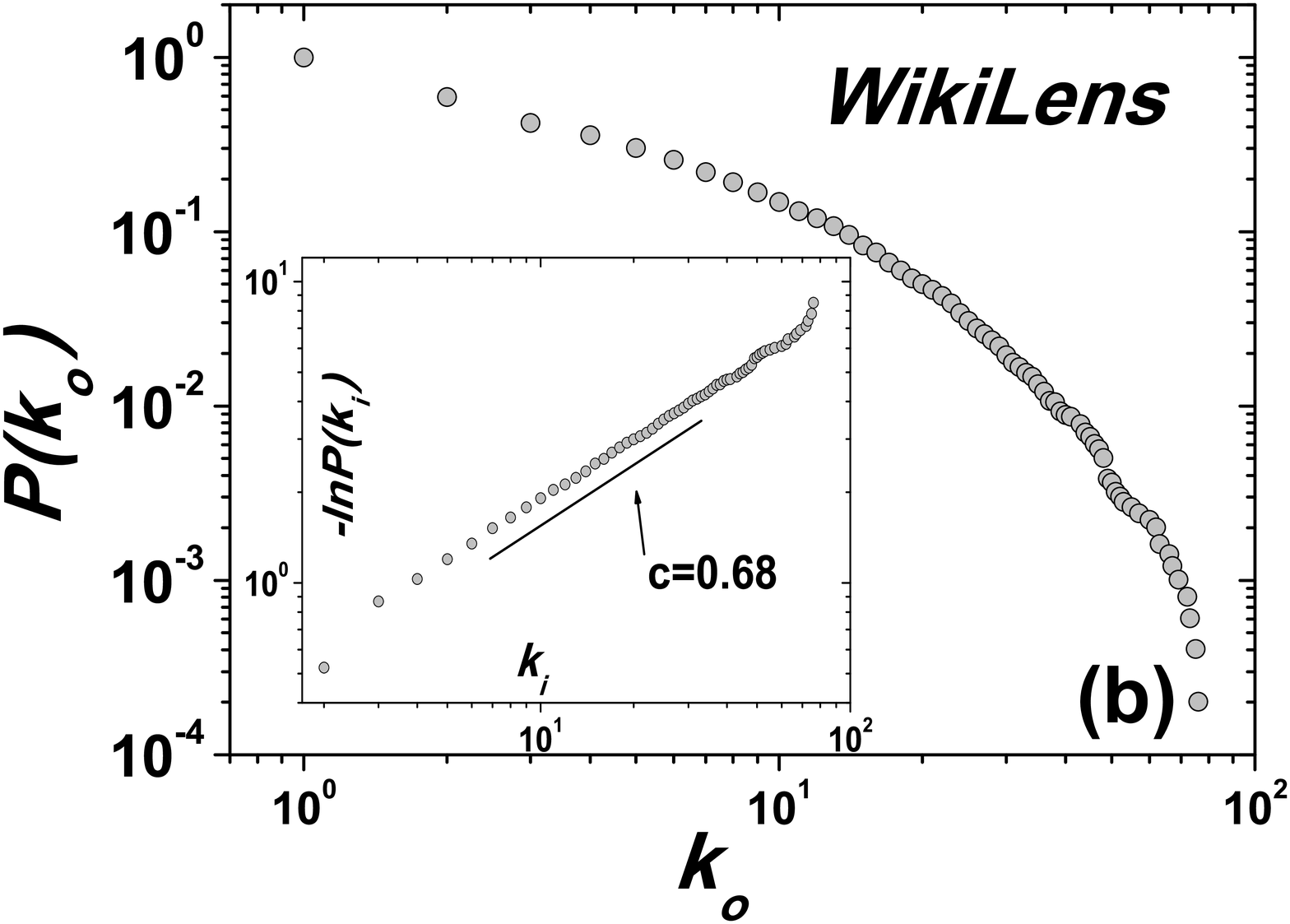}
\caption{ Distributions of object degrees, which also obey the stretched exponential form.}
\end{figure}

To demonstrate the presence of anchoring bias, we first look at an extreme case: Will we vote with systematic bias after voting on some very high-quality or very low-quality object--we name these objects as outliers. In the absence of systematic bias, the next votes after outliers' votes should be more or less the same to usual votes; while if the anchoring bias exists, a vote on an outlier will become the anchor of the next vote, and thus in average we will give high rating after voting on a high-quality object and low rating after a low-quality object.

\begin{table}[htbp]
\centering
\caption{\label{Table II}Basic statistics of outliers. The six columns from the second to the last one are the number of low-quality outliers (\#LQO), the number of votes right after votes on low-quality outliers (\#$A^{-}$), the number of high-quality outliers (\#HQO), the number of votes right after votes on high-quality outliers (\#$A^{+}$), the number of votes after votes on outliers (\#$A^{-}$\&$A^{+}$), and the percentage of these after-outlier votes in all votes. }
\begin{tabular}{cccccccc}
\hline
\hline
Data Set & \#LQO & \#$A^{-}$ & \#HQO & \#$A^{+}$ & \#$A^{-}$\&$A^{+}$ & Percentage\\
\hline
MovieLens & 97 & 8526 & 14 & 10713 & 19239 & 1.92\%  \\
WikiLens & 12 & 268 & 13 & 370 & 638 & 2.37\% \\
\hline
\hline
\end{tabular}
\end{table}

We use the average rating to estimate an object's quality, and to reduce the possible errors caused by personalized tastes and unreasonable votes, we only consider the objects getting more than ten votes. Although ratings cannot perfectly reflect qualities, they are correlated with qualities and can be naturally treated as anchors by users. For both MovieLens and WikiLens, an object (with more than ten votes) is distinguished as low-quality or high-quality outlier if its average rating is lower than 2.0 or higher than 4.5.

Denote by $r_{i\alpha}$ the rating from user $i$ onto object $\alpha$, and for an arbitrary user $i$, all her $k_i$ ratings are ordered by time as $r_{iO_{1}},r_{iO_{2}},r_{iO_{3}},\cdots,r_{iO_{k_{i}}}$, where $O_{1},O_{2},O_{3},\cdots,O_{k_{i}}$ are the objects having been voted by $i$, $r_{iO_{1}}$ is the oldest rating, and $r_{iO_{k_{i}}}$ is the most recent rating. If $O_{l} (l<k_{i})$ is a low-quality outlier, $r_{iO_{l+1}}$ is an after-low-quality-outlier rating ($A^{-}$ rating for short), while if $O_{l}$ is a high-quality outlier, $r_{iO_{l+1}}$ is an $A^{+}$ rating. According to the above criterion and definition, as shown in Table II, there are in total 19239 ($1.92\%$ of all ratings) after-outlier ratings for MovieLens and 638 ($2.37\%$ of all ratings) after-outlier ratings for WikiLens. One could observe that the high-quality outliers get more votes in average, which is in accordance with our common sense that better objects are more popular.

\begin{table}[htbp]
\centering
\caption{\label{Table III}Statistics of votes after outliers for MovieLens. $\langle r \rangle$, $\langle d_{o} \rangle$ and $\langle d_{u} \rangle$ respectively denote the average rating, the average difference to object average and the average difference to user average.}
\begin{tabular}{cccc}
\hline
\hline
$MovieLens$ & $\langle r \rangle$ & $\langle d_{o} \rangle$  & $\langle d_{u} \rangle$ \\
\hline
$A^{-} votes$ & 2.72 & -0.054 & -0.635 \\
$A^{+} votes$ & 4.16 & 0.033 & 0.449 \\
\hline
\hline
\end{tabular}
\end{table}

\begin{table}[htbp]
\centering
\caption{\label{Table IV}Statistics of votes after outliers for WikiLens. $\langle r \rangle$, $\langle d_{o} \rangle$ and $\langle d_{u} \rangle$ respectively denote the average rating, the average difference to object average and the average difference to user average.}
\begin{tabular}{cccc}
\hline
\hline
$WikiLens$ & $\langle r \rangle$ & $\langle d_{o} \rangle$  & $\langle d_{u} \rangle$ \\
\hline
$A^{-} votes$ & 2.63 & -0.071 & -0.850 \\
$A^{+} votes$ & 4.13 & 0.061 & 0.372 \\
\hline
\hline
\end{tabular}
\end{table}

We next compare the votes after low-quality outliers and those after high-quality outliers, namely $A^{-}$ and $A^{+}$ votes. The average rating among all $A^-$ votes is defined as
\begin{equation}
\langle r^{-} \rangle=\frac{1}{|A^{-}|}\sum_{r_{i\alpha}\in A^{-}}r_{i\alpha}.
\end{equation}
In addition, we look at the difference between a rating $r_{i\alpha}\in A^{-}$ and the average rating on the object $\alpha$, as well as the difference between $r_{i\alpha}$ and the average rating by the user $i$. Accordingly, we define the average difference to object average as
\begin{equation}
\langle d_{o}^{-} \rangle=\frac{1}{|A^{-}|}\sum_{r_{i\alpha}\in A^{-}}(r_{i\alpha}-\langle r_{\bullet \alpha} \rangle),
\end{equation}
where $\langle r_{\bullet \alpha} \rangle$ denotes the average rating on $\alpha$, and the average difference to user average as
\begin{equation}
\langle d_{u}^{-} \rangle=\frac{1}{|A^{-}|}\sum_{r_{i\alpha}\in A^{-}}(r_{i\alpha}-\langle r_{i\bullet} \rangle),
\end{equation}
where $\langle r_{i\bullet} \rangle$ denotes the average rating by $i$. Analogously, we can define $\langle r^{+} \rangle$, $\langle d_{o}^{+} \rangle$ and $\langle d_{u}^{+} \rangle$ for $A^{+}$ votes.

Table III and Table IV show the remarkable difference between people's votes after low-quality and high-quality outliers, respectively. The results indicate the possible presence of anchoring bias, that is, people tend to give a low rating if the prior-visited object is not good, and vice versa. However, the above evidence is not solid enough since it covers only a tiny fraction of votes, and thus we will further analyze the full rating series of every user.

To get rid of the effects of different voting standards of users (e.g., some users are good-tempered and tend to give high ratings than others) and different deserving ratings of objects (e.g., some objects are of better qualities and should be voted with high ratings), we regulate each rating $r_{i\alpha}$ in the following four ways: (i) to eliminate the average rating over all votes as $r_{i\alpha}'=r_{i\alpha}-\langle r \rangle$; (ii) to eliminate the average rating over all votes on the corresponding object as $r_{i\alpha}'=r_{i\alpha}-\langle r_{\bullet \alpha} \rangle$; (iii) to eliminate the average rating over all votes on the corresponding user as $r_{i\alpha}'=r_{i\alpha}-\langle r_{i\bullet} \rangle$; (iv) to eliminate both average ratings as $r_{i\alpha}'=(r_{i\alpha}-\langle r_{\bullet \alpha} \rangle)+(r_{i\alpha}-\langle r_{i\bullet} \rangle)$. Readers are easy to reproduce all the following experiments and will find that the four cases lead to qualitatively the same results, and thus we only present the results of case (iv) hereinafter and without specific statement, the term rating(s) stands for regulated rating(s) of case (iv).

\begin{figure}
\includegraphics[width=3.5in]{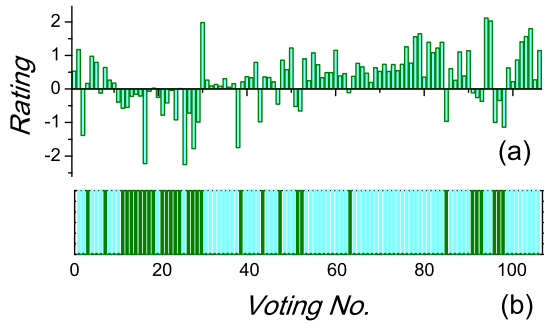}
\caption{(Color online) Rating series of a typical user in MovieLens who has voted 106 movies in total. All these 106 ratings are displayed according to the voting time in panel (a), and the positive and negative ratings are respectively represented by light-green and dark-green lines in panel (b).}
\end{figure}

Figure 3(a) presents the rating series of a typical user in MovieLens. We simply divide ratings into two classes--positive ratings and negative ratings, and display them without explicit values in figure 3(b), where one could observe that ratings in the same class are aggregated. This kind of aggregation reveals the anchoring bias in voting behavior, namely people is likely to give a high rating after a prior high rating while is likely to give a low rating after a prior low rating. Similar to the method used to measure the memory effect of a time series \cite{Goh2008}, to quantify the aggregation phenomenon for an arbitrary user $i$, we calculate the Pearson correlation coefficient $R_{i} (-1\leq R_{i} \leq 1)$ of two series $r_{iO_{1}}$, $r_{iO_{2}}$, $\cdots$,$r_{iO_{k_{i}-1}}$  and $r_{iO_{2}}$, $r_{iO_{3}}$, $\cdots$,$r_{iO_{k_{i}}}$, where the Pearson correlation coefficient for two finite series $x_{1}, x_{2}, \cdots, x_{n}$ and $y_{1}, y_{2}, \cdots, y_{n}$ is defined as
\begin{equation}
R(x,y)=\frac{\sum_{i=1}^{n}(x_{i}-\langle x \rangle)(y_{i}-\langle y \rangle)}{\sqrt{\sum_{i=1}^{n}(x_{i}-\langle x \rangle)^{2}}\sqrt{\sum_{i=1}^{n}(y_{i}-\langle y \rangle)^{2}}}.
\end{equation}
According to the definition, a positive $R_{i}$ indicates that the user $i$ may have the anchoring bias.

\begin{figure}
\includegraphics[width=3.5in]{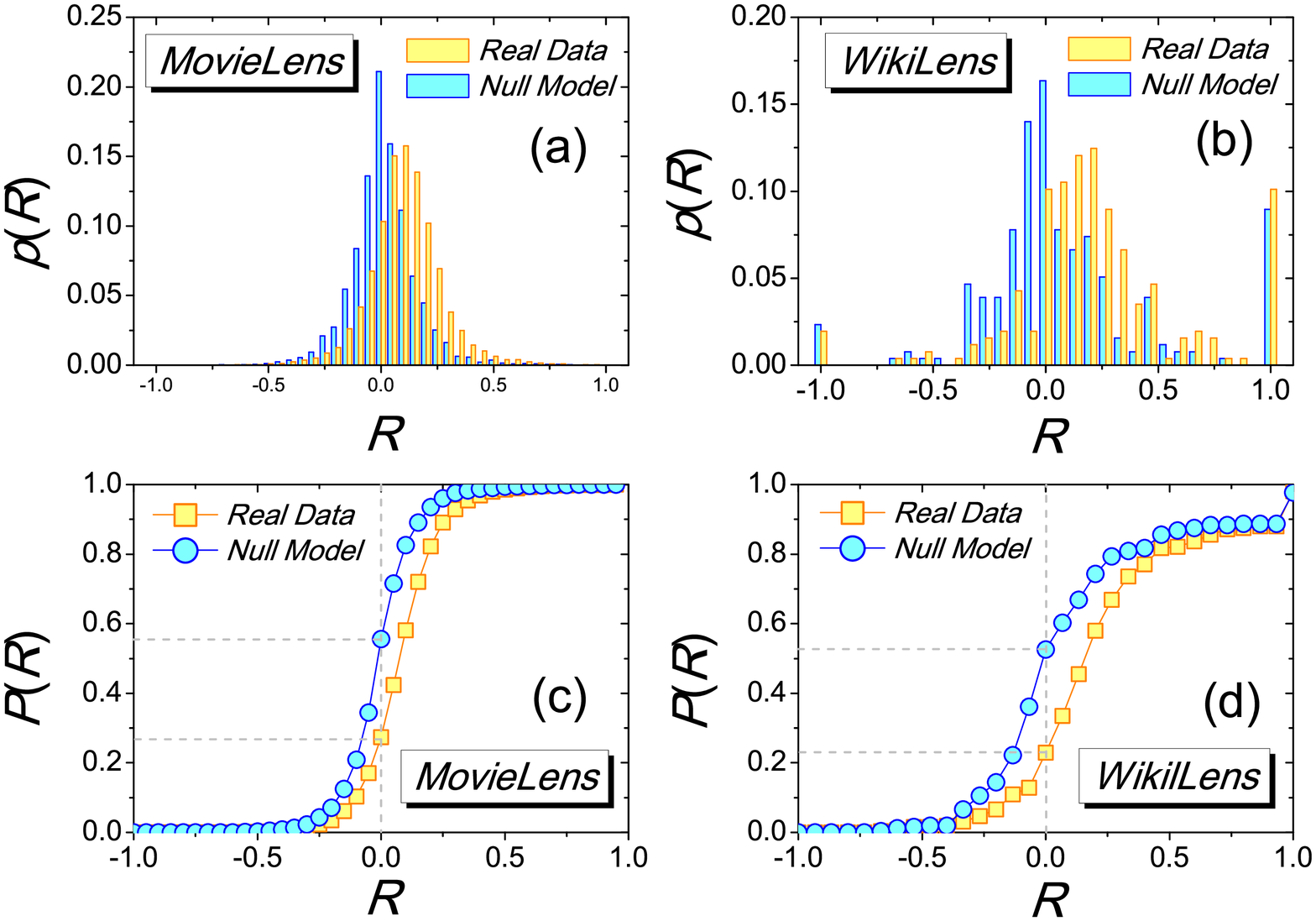}
\caption{(Color online) Distributions of users' Pearson correlation coefficients $R$. (a) and (c) are for MovieLens, while (b) and (d) are for WikiLens. (a) and (b) show the histograms where $p(R)$ is the probability density of $R$, while (c) and (d) show the cumulative distributions where $P(R)$ denotes the fraction of users whose Pearson correlation coefficients are less than $R$. In each panel, results from the real data and the null model are respectively colored in red and blue. }
\end{figure}

\begin{figure}
\includegraphics[width=3.5in]{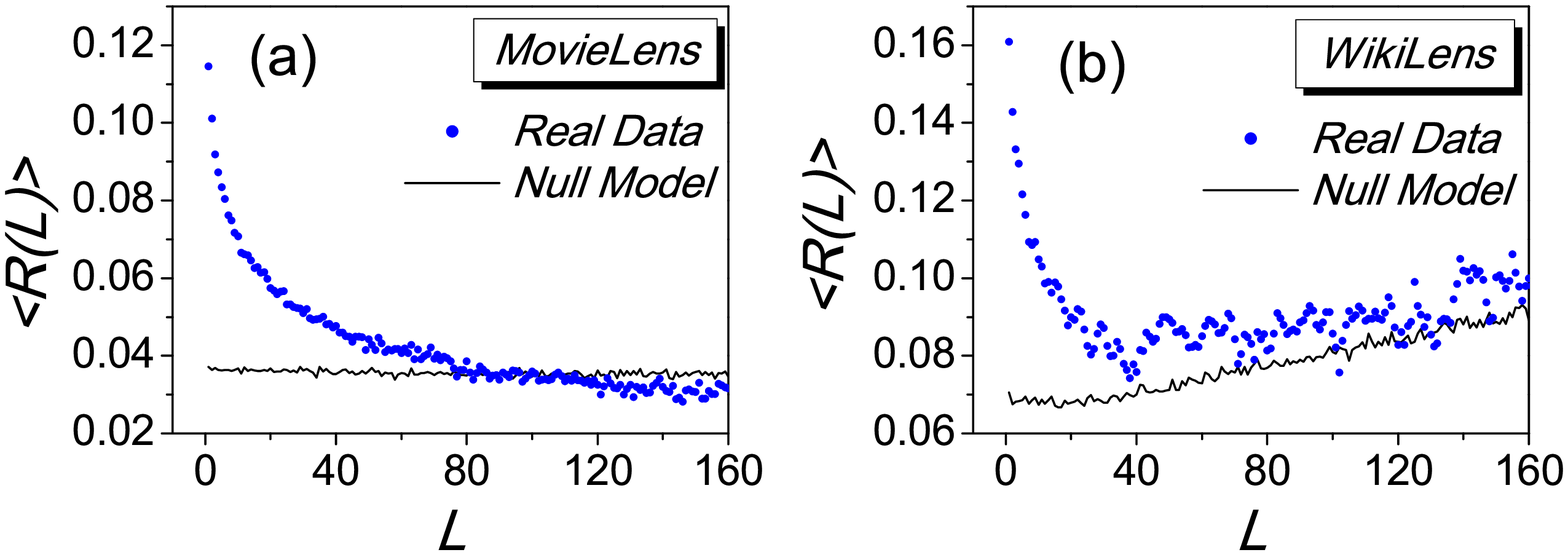}
\includegraphics[width=3.5in]{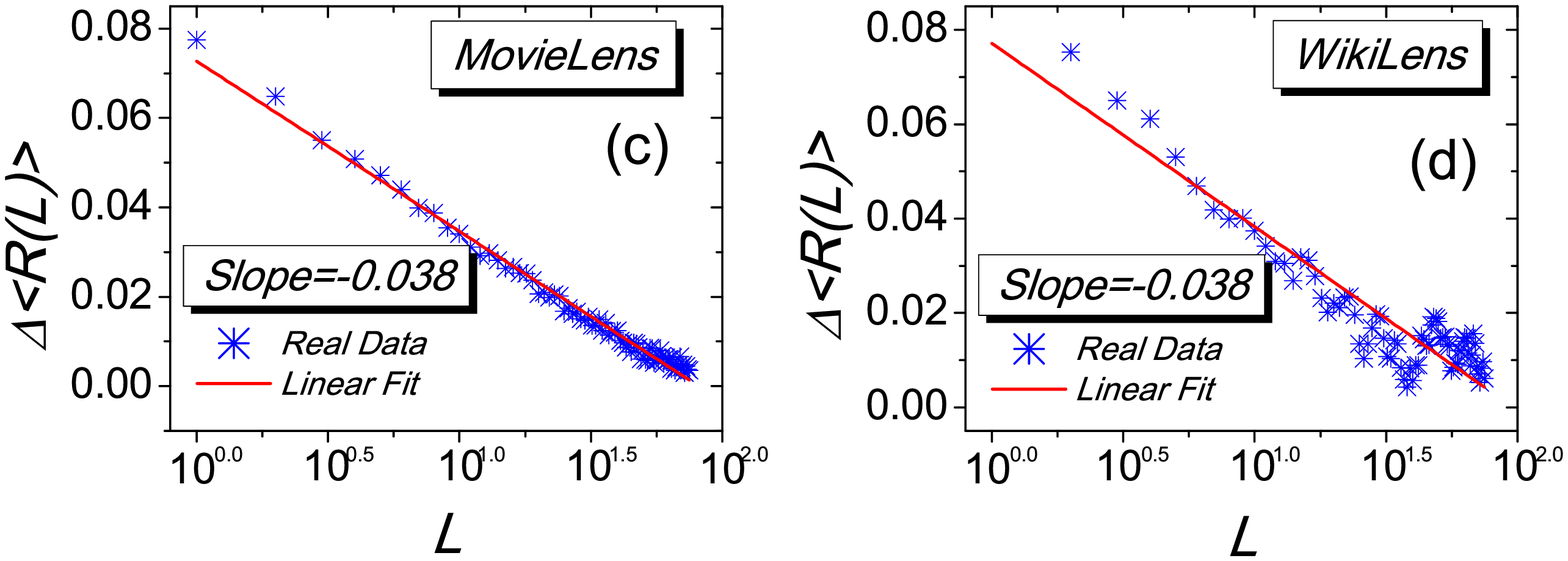}
\caption{(Color online) The average Pearson correlation coefficient $R(L)$ versus the correlation length $L$ for MovieLens (a) and WikiLens (b). The blue and black curves respectively represent the results of the real data and the null model. Panels (c) and (d) display how the difference between $R(L)$ of the real data and the null model changes with $L$ in the linear-log scale.}
\end{figure}

We compare the empirical results with those of a null model, in which each user's voting times are randomly redistributed. That is to say, for an arbitrary user $i$, the rating series $\{r_{iO_{1}}$, $r_{iO_{2}}$, $\cdots$,$r_{iO_{k_{i}}}\}$, is reordered. Figure 4 compares the distributions of Pearson correlation coefficients $R$ of the empirical data and the null model. Clearly, the distributions of the null model peak at about zero, while the empirical distributions peak at a positive value. In addition, the empirical distributions, as a whole, lie in the right of the distributions of the null model. From the cumulative distributions, one could see that for the null model, less than $50\%$ of users are of positive $R$, while for the empirical data, more than $70\%$ of users are of positive $R$. Aforementioned comparison shows the significance of the anchoring bias in empirical data.

If the number of votes of a user $i$ is larger than $L$, we could extend the Pearson correlation coefficient $R_{i}$ to an $L$-dependent coefficient $R_{i}(L)$ as the Pearson correlation coefficient of two series $r_{iO_{1}}$, $r_{iO_{2}}$, $\cdots$,$r_{iO_{k_{i}-L}}$ and $r_{iO_{L+1}}$, $r_{iO_{L+2}}$, $\cdots$,$r_{iO_{k_{i}}}$. As shown in figure 5, the average value of the $L$-dependent Pearson correlation coefficient, $\langle R(L) \rangle$, of the empirical data over all users is remarkably larger than that of the null model for small $L$, and the difference between $\langle R(L) \rangle$ of the empirical data and the null model decays in a logarithmic way as
\begin{equation}
\Delta \langle R(L) \rangle \approx A-B \log L,
\end{equation}
where $A \approx 0.08$ and $B \approx 0.04$ for both MovieLens and WikiLens. This result again indicates the existence of the anchoring bias. Moreover, it suggests that this bias will last a considerable time period, which is in accordance with previous experimental results on the duration of anchoring effects \cite{Mussweiler2001}.

Combining those aforementioned experiments, the existence and significance of the anchoring bias in online voting is obviously validated, whose pattern, as shown in figure 3, is very similar to the memory-embedded time series \cite{Goh2008}. The extent of the anchoring bias, quantified by the difference of the average regulated rating from the null model, decays in a logarithmic form and will last a considerable duration. Most known literature on anchoring effects considered people's judgements, evaluations, estimations and predictions in offline world, meanwhile the quick development of Internet and the data processing technologies allow us to study the rich social psychological phenomena in online world. Quantitative analysis and statistical description based on BIG DATA may build up a new paradigm for social psychology and facilitate the birth of a new branch of psychology, probably called \emph{Internet psychology}. This work is an elementary attempt that tries to uncover underlying decision-making processes based on extensive statistical analysis. Our findings are helpful in understanding the online voting pattern and improving the performance of recommender systems.

\acknowledgements

This work is partially supported by the Fundamental Research Funds for the Central Universities of China and the National Natural Science Foundation of China under Grant Nos. 11222543 and 11105024.

\end{document}